\documentclass[paper]{JHEP3}

\usepackage{epsfig,multicol,bbm}

\def\){\right)}
\def\({\left( }
\def\]{\right] }
\def\[{\left[ }

\def\no{\nonumber \\}

\def\be{\begin{equation}}
\def\ee{\end{equation}}
\def\ba{\begin{eqnarray}}
\def\ea{\end{eqnarray}}
\def\no{\nonumber \\}

\newcommand\fverb{\setbox\fverbbox=\hbox\bgroup\verb}
\newcommand\fverbdo{\egroup\medskip\noindent%
			\fbox{\unhbox\fverbbox}\ }
\newcommand\fverbit{\egroup\item[\fbox{\unhbox\fverbbox}]}
\newbox\fverbbox

\title{K\"{a}hler moduli inflation and WMAP7
 }
\author{Sunggeun Lee and Soonkeon Nam \\
	Department of Physics and Research Institute for Basic Sciences, \\
   Kyung Hee University, Seoul 130-701, Korea\\
	E-mail: \email{nam@khu.ac.kr}}

\abstract{Inflationary potentials are investigated for specific
models in type IIB string theory via flux compactification. As concrete models, we investigate several cases where the internal spaces are weighted projective spaces. The models we consider have two, three, or four K\"{a}hler moduli. The K\"{a}hler moduli play a role of inflaton fields and we consider the cases where only one of the moduli behaves as the inflaton field. For the cases with more than two moduli, we choose the diagonal basis for the expression of the Calabi-Yau volume, which can be written down as a function of four-cycle. With the combination of multiple moduli, we can express the multi-dimensional problem as an effective one-dimensional problem. In the large volume scenario, the potentials of these three models turn out to be of the same type. By taking the specific limit of the relation between the moduli and the volume, the potentials are reduced to simpler ones which induce inflation. As a toy model we first consider the simple potential. We calculate the slow roll parameters $\epsilon$, $\eta$ and $\xi$ for each inflationary potential. Then, we check whether the potentials give reasonable spectral indices $n_s$ and their running $\alpha_s$'s by comparing with the recently released seven-year WMAP data. For both models, we see reasonable spectral indices for the number of e-folding $47<N_e<61$. Conversely, by inserting the observed seven-year WMAP data, we see that the potential of the toy model gives requisite number of e-folds while the potential of the K\"{a}hler moduli gives much smaller number of e-folding. Finally, we see that two models do not produce reasonable values of the running of the spectral index.}

\begin{document}

\section{Introduction}
\setcounter{equation}{0}
What has been a big challenge in theoretical physics in recent decades is to explain our present universe. It has been observed that the universe is expanding, especially with acceleration. Thus, the accepted standard big bang theory should be supplemented by inflationary period \cite{linde1,kolb} to extrapolate from the observed current universe to the early universe. Therefore, the inflation has been one of the most important paradigm in modern cosmology. The inflationary model was first introduced in Ref.\cite{Guth} and developed by many others \cite{olive}. In the absence of quantum gravity, these models are effective field theories minimally coupled to scalar fields with Einstein gravity. The problem is that the choice of inflaton and its potential is rather ad hoc. Solving this problem requires studying some fundamental theory such as string theory. It seems very reasonable to investigate the inflationary model from the point of a fundamental theory such as string theory. In this top down approach, we have to find the inflaton field and have a nice potential such that the slow roll parameters $\epsilon$, $\eta$, and $\xi$ are small enough to allow enough number of e-folding. In general, it is complicated to work out the potential. However, luckily, in type IIB string theory, several examples including non-perturbative potential were worked out in great detail.

The string theory, which is the leading candidate for the quantum gravity, has been an
active research area recently for the application to cosmology \cite{quevedo,mcasilv}. However, the four dimensional theory compactified on six dimensional Calabi-Yau manifold accompanies lots of massless fields called moduli. Recent study resolves this problem by turning on fluxes appropriately \cite{gidd}. However, by turning on the fluxes alone, the vacuum of the effective four dimensional theory just becomes supersymmetric anti de Sitter (AdS) spacetime. Based on this approach, the meta-stable de Sitter vacua was constructed by Kachru, Kallosh, Linde and Trivedi (called KKLT scenario) in Ref.\cite{kklt}. They obtained de Sitter vacuum by breaking supersymmetry with the addition of anti-brane and by including a uplift term. Both of these effects uplift the negative vacuum energy to positive vacuum energy. Along the KKLT scenario, the application to inflationary cosmology was triggered by the work of Kachru et al.\cite{kklmmt}.
Even though we have the quite successful de Sitter vacuum, the effective four-dimensional theory compactified on Calabi-Yau space may give rise to many degenerate vacua often called landscape. In other words, different inflationary scenarios can be realized in different regions of the string theory landscape. Therefore, finding the de Sitter vacua is not unique, and there can be many attempts to do it.

In string theory there are various moduli that can be interpreted as inflaton: the first one is the distance between the branes (D-brane inflation \cite{dvtye}) or between brane and anti-brane \cite{burg}. In this case the effective mass is often too big to give enough e-folds.
The second one is the one from geometric moduli such as K\"{a}hler moduli which are the four-cycle inside the Calabi-Yau. In Ref.\cite{conquev}, the K\"{a}hler moduli inflation in type IIB string theory was studied based on so called large volume scenario \cite{balasubb}. It is required that at least two moduli be needed for the inflationary model. The role of axionic field in K\"{a}hler inflation was studied in Ref.\cite{roulett}.
Any realistic string theoretic model of the inflation, formulated in terms of an effective four dimensional supergravity, must have all moduli (axion-dilaton, complex structure, K\"{a}hler moduli, and brane positions) stabilized, and have at least one inflaton, with a potential flat enough to provide a slow-roll evolution. There are other models in string theory discussing cosmological scenarios: brane gas model \cite{Battefeld}, pre-big bang scenario \cite{gasven}, rolling tachyon \cite{sen}, ekpyrotic \cite{khour}, D-term \cite{bd,kl}, and racetrack inflation \cite{blanco}.

The reason we focus on the K\"{a}hler moduli is that the potential has almost flat direction and both moduli stabilization and flatness of the potential are achieved by the same mechanism. In the case of the large volume scenario of the flux compactifications, these are valid for a very large class of models. The problem in studying models with multiple moduli is that the field space is multi-dimensional, and the trajectory of the inflaton field during inflation can be quite complicated. In the particular model with four moduli, we study ${\bf P}^4_{1,2,2,10,15}$ model \cite{coll}. This model is one of Swiss-cheese models. The volume of the Calabi-Yau manifold is determined by the biggest cycle of the manifold and the small four-cycles make holes inside of Calabi-Yau manifold: the volume of the Calabi-Yau manifold has a form like ${\cal V}=T_L^{3\over 2}-\sum_s T_s^{3\over 2}$ where $T$'s are four-cycles. We deal with the multi-dimensional problem by choosing the diagonal basis and then pick one of the moduli as an inflaton.

This paper is organized as follows: In section 2, we review flux compactification in type IIB string theory including the large volume scenario and also discuss the string one loop correction to the potnetial. The flux stabilizes the axion field, the complex structure, and the dilaton moduli. The non-perturbative effect stabilizes K\"{a}hler structure moduli. In section 3, the derivation of the inflationary potential from large volume scenario is reviewed. The potential obtained in large volume scenario is reduced to simpler one when the K\"{a}hler moduli has a special limit. In section 4, we apply the paradigm of section 3 to Swiss-cheese models with two, three, or four K\"{a}hler moduli. We first consider the simplest model, $P^4_{1,1,1,6,9}$, which has two K\"{a}hler moduli. Next, we study two models with three moduli, Fano $({\cal F}_{11})$ \cite{denefdoug} and ${\bf P}^4_{1,3,3,3,5}$ \cite{blum}. Finally, the model with four moduli, ${\bf P}^4_{1,2,2,10,15}$ \cite{coll}, is investigated. For more than two moduli, we take the diagonal basis for the volume of the the Calabi-Yau manifold. This makes it simpler to handle the analysis for finding the inflaton field and studying the inflation. When we consider the inflationary potential, all the above models are reduced to the same form of potential. In section 5, we study two cases as  inflationary models. In the first case, we study the inflationary potential starting with a toy model. In the second case, we study K\"{a}hler moduli inflation with canonically normalized inflaton field. We calculate the number of e-folding, $N_e$, and slow roll parameters, $\epsilon$, $\eta$, and $\xi$, for each model which make it possible to find the spectral index and its running. For both cases, we find that the slow roll parameters can be expressed as a number of e-folding. For the values of the allowed number of e-folding $(47<N_e<61)$, the slow roll parameters have reasonable values. Then, we compare the spectral index and its running with the seven-year WMAP data. We see that the spectral indices for both cases fit the seven-year WMAP data while the results of the running do not fit the seven-year WMAP data. We conclude with the discussion.

\section{A short review of the type IIB flux compactification}
\setcounter{equation}{0}
When we compactify the ten-dimensional string theory on real six-dimensional Calabi-Yau manifold, the effective four-dimensional theory has lots of massless fields called moduli. In nature, however, since these moduli fields have not been detected observationally, we need to resolve this problem. One way to give them masses is to induce the potential. In recent study \cite{gidd}, it was shown that the moduli fields can have masses by turning on fluxes. In the present work, we will work in type IIB string theory compactified on Calabi-Yau orientifolds, with RR and NS-NS 3-form fluxes by $F_3$ and $H_3$, respectively \cite{gidd}. See \cite{doug,grana,denef} for a review. The 3-form fluxes are quantized as
\be
{1\over {(2\pi)^2\alpha'}}\int_{\Sigma_a}F_3=n_a \in Z,~~~{1\over {(2\pi)^2\alpha'}}\int_{\Sigma_b}H_3=m_b\in Z,
\ee
where $\Sigma_a$ and $\Sigma_b$ are 3-cycles in the internal Calabi-Yau space $M$ and $\alpha'$ is related to string length $l_s$ such that $\alpha'=l_s^2/4\pi^2$.
Ignoring the gauge sectors, the theory is specified by the K\"{a}hler potential $K$ and superpotential $W$. The superpotential $W$ does not depend on the K\"{a}hler moduli and takes the form
\be
W=\int_M G_3 \wedge \Omega,
\ee
where 3-form $G_3$ is given by $G_3=F_3+iS H_3$. Here, the axion-dilaton field $S$ is given by $S=e^{-\phi}+i C_0$ where $\phi$ and $C_0$ are dilaton and RR 0-from, respectively. The $\Omega$ is the holomorphic (3,0) form of the internal Calabi-Yau manifold.
The K\"{a}hler potential up to leading order in string coupling $g_s$ and $\alpha'$ is given by
\be
K\equiv {K_{no-scale}\over M_p^2}=-2\log [{\cal V}_s]-\log \[ -i \int_M\Omega\wedge {\bar \Omega}\] -\log[S+\bar S],
\ee
where the volume of Calabi-Yau, ${\cal V}$, is given by ${\cal V}={\cal V}_s l_s^6$. In $N=1$ supergravity, the scalar potential is known as
\be
V=  e^K\[K^{i\bar{j}}D_iW{\bar D}_j{\bar W}-3|W|^2\] ,
\ee
where the K\"{a}hler metric $K_{i\bar j}$ is defined by $K_{i\bar j}=\partial_i  \partial_{\bar j} K$ and $i,j$ run over all moduli. Moreover, the derivative $D_i$ is defined by $D_i W=\partial_i W +\partial_i K W$. The superpotential $W$ is independent of the K\"{a}hler miduli, and the term $-3|W|^2$ is canceled \cite{gidd}. Hence, this scalar potential is reduced to
\be
V_{no-scale}= e^K K^{a{\bar b}}D_aW {\bar D}_b{\bar W},\label{sugrapot}
\ee
where $a$ and $b$ run over dilaton and complex structure moduli. The minimum of this potential is found by solving
\be
D_aW=0.
\ee

The $\alpha'$ correction to K\"{a}hler potential was obtained in \cite{bbhl} :
\be
K_{\alpha'}=-2\log\[{\cal V}_s +{\xi \over {2 g_s^{3\over 2}}}\]-\log\[-i\int_M\Omega\wedge\bar \Omega\]-\log[S+\bar S],
\ee
where $\xi=-{\zeta(3) \chi(M)\over {2(2\pi)^3}}$ and $\chi(M)$ is Euler number of $M$. We require $\xi>0$ and the hodge numbers $h^{2,1}$ and $h^{1,1}$ should have the relation such that $h^{2,1}>h^{1,1}$.

The superpotential receives no $\alpha'$ corrections. However, it receives
non-perturbative correction of K\"{a}hler moduli through D3-brane instantons or gaugino condensation
from wrapped D7-branes. It takes the form
\be
W=\int_MG_3\wedge\Omega+\sum_iA_ie^{-a_iT_i},
\ee
where $A_i$ is a one-loop determinant. The K\"{a}hler moduli $T_i$ is defined by $T_i=\tau_i+ib_i$ where $\tau_i$ is the volume of a 4-cycle in $M$, and $b_i$ is given by $b_i=\int C_4$ where $C_4$ is RR four-form. For D3-brane instanton, $A_i$ only depends on the complex structure moduli and $a_i={2\pi \over N}$ with $N\in Z_+$ and $N=1$ for D3-instantons. In the following, we consider $a_i=2\pi$.

After fixing or integrating out the dilaton and complex structure moduli, the K\"{a}hler potential becomes
\ba
K&=&K_{cs}-2\log\[{\cal V}_s +{\xi \over 2}\],\no
W&=&W_0+\sum_iA_ie^{-a_iT_i}\label{kahsuppot}.
\ea
Here, $W_0={\hat W}_0l_s^4$ and $A_n={\hat A}_n l_s^4$ where ${\hat W}_0$ and ${\hat A}_n$ are dimensionless numbers.
Substituting two equations in (\ref{kahsuppot}) into (\ref{sugrapot}), we get
\ba
&&V=e^K\[K^{T_j\bar T_k}\(a_jA_ja_k\bar A_ke^{-(a_jT_j-a_k\bar T_k)}+i(a_jA_je^{-a_jT_j}\bar W\partial_{\bar T_k}K-a_k\bar A_k e^{-a_k\bar T_k}W\partial_{T_j}K)\)\right.\no
&&~~~~~\left.+3\xi {{(\xi^2+7\xi {\cal V}_s +{\cal V}_s^2)}\over {({\cal V}_s-\xi)(2{\cal V}_s+\xi)^2}}|W|^2\]\no
&&\equiv V_{np1}+V_{np2}+V_{\alpha'}\label{potential}.
\ea
There are two stages for large volume AdS minimum of the potential (\ref{potential}):
\begin{itemize}
\item
The four-cycle volumes $\tau_i\equiv Re(T_i)\to \infty$.
\item
$V<0$ for large ${\cal V}_s$ and the potential approaches zero from below.
\end{itemize}
At large volume ${\cal V}_s\to \infty$, the $\alpha'$ correction term $V_{\alpha'}$ proportional to ${1\over {\cal V}_s^3}$ will dominate over the non-perturbative terms $V_{np1}$ and $V_{np2}$. The non-perturbative terms which are exponentially suppressed can compete with perturbative terms.
When we consider the limit ${\cal V}_s\to \infty$, the moduli $\tau_i$ are taken to have a limit $\tau_i\to \infty$ except the smallest one denoted by $\tau_s$ with the relation $a_s\tau_s=\ln {\cal V}_s$.
As a result we get the potential in the large volume limit:
\be
V\sim \[ {1\over {\cal V}_s}a_s^2 |A_s|^2 (-k_{ssk}t^k)e^{-2a_s\tau_s}-{1\over {\cal V}_s^2}a_s\tau_se^{-a_s\tau_s}|A_sW|+{\xi \over {\cal V}_s^3}|W|^2\],
\ee
where $k_{ssk}$ is an intersection number and $t^k$ are two-cycle volumes of the Calabi-Yau manifold.

In \cite{bhp,cicolia,cicob}, the stringy one loop correction in string coupling $g_s$ was considered.
There are two types of contributions to the K\"{a}hler potential, coming from winding mode, $\delta K^W$, and Kaluza-Klein mode, $\delta K^{KK}$. The correction term to the potential can be written as
\be
\delta V^{1-loop}_{(g_s)}=\sum^{h_{1,1}}_{i=1}\( {({\cal C}^{KK}_i)^2\over {Re(S)^2}}(-2\partial_i\partial_i \ln {\cal V}_s)-2{ {\cal C}^W_i\over {(a_{il}t^l){\cal V}_s}}\){W^2_0\over {{\cal V}_s}},
\ee
where ${\cal C}^{KK}_i$ and ${\cal C}^W_i$ are unknown functions on the complex structure moduli. After stabilizing the complex structure moduli, these become unknown constants. This one loop correction plays an important role in stabilizing the moduli for the cases with three K\"{a}hler moduli in the following section.

\section{The K\"{a}hler moduli inflation}
\setcounter{equation}{0}
In this section, we consider inflationary model with the potential arising from the flux compactification of type IIB string theory. The various K\"{a}hler moduli play a role of inflaton field. We focus on the model that only one of K\"{a}hler moduli plays as an inflaton field. During the inflation, the size of the bulk six dimensional Calabi-Yau volume is supposed to be fixed while the variation of the internal small four-cycle induces an inflation.
We are assuming that the inflaton field rolls from a point which is far from the the stable point. Inflation will ends when the inflaton reaches the stable point where the moduli get fixed.

In the following, we will consider the model of which the volume of the Calabi-Yau can be calculated.
First, as an illustration, we will consider simplified Calabi-Yau volume,
\be
{\cal V}_s=\alpha\( \tau^{3\over 2}_1-\sum^n_{i=2}\lambda_i\tau^{3\over 2}_i\)
={\alpha\over {2\sqrt{2}}}\[(T_1+\bar T_1)^{3\over 2}-\sum^n_{i=1}\lambda_i(T_i+\bar T_i)^{3\over 2}\],\label{volume}
\ee
where $\tau_1$ controls the overall volume and $\tau_2,\cdots,\tau_n$ are blow-ups whose only non-vanishing triple intersections are with themselves. $\alpha$ and $\lambda_i$ are positive constants and depend on particular model. Later, we consider the model with the number of moduli fields, two, three, or four. Then, we see that by diagonalization of the volume the three models arrive at the form of Eq.(\ref{volume}).

The dilaton and complex structure moduli are stabilized by fluxes and the K\"{a}hler moduli are stabilized by superpotential $W$ in Eq. (\ref{kahsuppot}) and the K\"{a}hler potential $K$ with $\alpha'$ correction is given by
\be
K=K_{cs}-2\log\[ \alpha\(\tau^{3\over 2}_1-\sum^n_{i=2}\lambda_i\tau^{3\over 2}_i\)+{\xi \over 2}\].
\ee
For large volume scenario $\tau_1\gg \tau_i$ and ${\cal V}_s\gg 1$, the scalar potential becomes \cite{conquev}
\be
V=\sum_i{8(a_i\hat A_i)^2\sqrt{\tau_i}\over {3{\cal V}_s \lambda_i\alpha}}e^{-2a_i\tau_i}-\sum_i4{a_i\hat A_i\over {\cal V}_s^2} \hat W_0\tau_ie^{-a_i\tau_i}+{3\xi \hat W_0^2\over {4{\cal V}_s^3}}.\label{pot1}
\ee
We will discuss the inflationary potential derived from the above potential in Eq.(\ref{pot1}). One of K\"{a}hler moduli $\tau_n$ will play as an inflaton field.
By taking the limit $e^{a_n\tau_n}\gg {\cal V}_s^2$, the above potential is simplified to
\be
V_{inf}=V_0-{4\tau_n\hat W_0a_n \hat A_ne^{-a_n\tau_n}\over {{\cal V}_s^2}}\label{infpot},
\ee
where the constant $V_0$ is $V_0={\beta \hat W_0^2\over {\cal V}_s^3}$ with $\beta={3\xi\over 4}$.

\begin{figure}
\begin{center}
\includegraphics[width=2.8in]{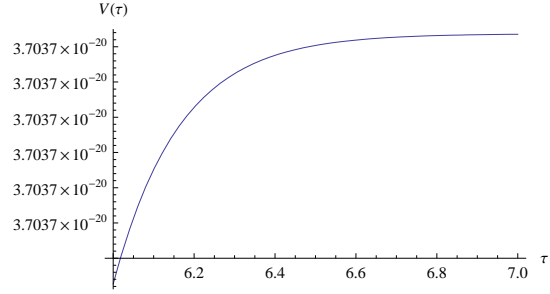}
\end{center}
\caption{Plot of the potential $V_{infl}$ with numerical values $\hat W_0=1, \beta=1, a_n=2\pi, \hat A_n=1, \lambda=1$.}\label{kpoten}
\end{figure}

The kinetic term couples to K\"{a}hler metric and we need to redefine the field to consider ordinary inflationary potential.
Then, the canonically normalized field is obtained through straightforward calculation:
\be
\tau^c_{n}=\sqrt{{4\lambda\over {3{\cal V}_s}}}\tau^{3\over 4}_n.
\ee
In terms of $\tau^c_{n}$, the inflationary potential becomes
\be
V_{infl}=V_0-{4\hat  W_0a_n \hat A_n\over {\cal V}_s^2}\({3{\cal V}_s\over {4\lambda}}\)^{2\over 3}(\tau_n^c)^{4\over 3} e^{-a_n\({3{\cal V}_s\over {4\lambda}}\)^{2\over 3}(\tau_n^c)^{4\over 3}}.\label{kpot}
\ee
The plot of this inflationary potential is shown in Fig. (\ref{kpoten}).
In order to see if the potential gives enough inflation, we have to compute the so called slow roll parameters. The slow roll parameters are defined by
\be
\epsilon={M_p^2\over 2}\({V'\over V}\)^2,~~~\eta=M_p^2{V''\over V},~~~
\xi=M_p^4{V'V'''\over V^2}.
\ee
By straightforward calculation with the given potential, we express the parameters as original K\"{a}hler moduli:
\ba
\epsilon&=&{32{\cal V}_s^3\over {3\beta^2W_0^2}}a_n^2A_n^2\sqrt{\tau_n}(1-a_n\tau_n)^2e^{-2a_n\tau_n},\no
\eta&=&-{4a_nA_n{\cal V}_s^2\over {3\lambda\sqrt{\tau_n}\beta W_0}}[(1-9a_n\tau_n+4(a_n\tau_n)^2)e^{-a_n\tau_n}],\no
\xi&=&-{32(a_nA_n)^2{\cal V}_s^4\over {9\beta^2\lambda^2W_0^2\tau_n}}(1-a_n\tau_n)(1+11a_n\tau_n-30(a_n\tau_n)^2+8(a_n\tau_n)^3)
e^{-2a_n\tau_n}.
\ea
The number of e-folding can be read with from the potential:
\be
N_e=\int^\phi_{\phi_{end}}{V\over V'}d\phi=-{3\beta W_0\lambda_n\over {16{\cal V}_s^2a_nA_n}}\int^{\tau_n}_{\tau^{end}_n}{e^{a_n\tau_n}\over {\sqrt{\tau_n}(1-a_n\tau_n)}}d\tau_n.
\ee

\begin{figure}
\begin{center}
\includegraphics[width=2.8in]{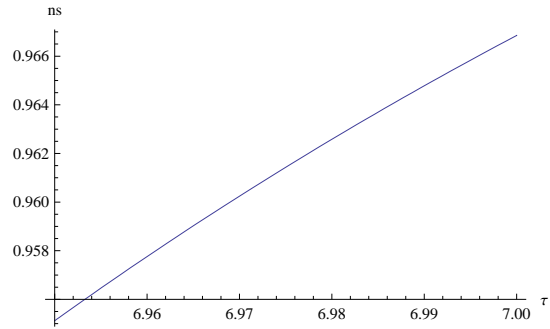}
\end{center}
\caption{The spectral index $n$ for $\hat W_0=1$, $\beta=1$, $a_n=2\pi$, $\hat A_n=1$, $\lambda=1$, with ${\cal V}_s=3\times 10^6$. $N_e=55$}\label{k123}
\end{figure}

Providing $e^{a_n\tau_n}\gg {\cal V}_s^2$, we may estimate each slow parameter:
In this limit the e-folding $N_e$ goes to
\be
N_e\simeq {3\beta W_0 \lambda_n\over {16a_nA_n{\cal V}_s^2\sqrt{a_n}}}{e^{a_n\tau_n}\over {(a_n\tau_n)^{3\over 2}}},\label{ne}
\ee
and the three slow parameters $\epsilon$, $\eta$, and $\xi$ goes to
\ba
&&\epsilon\simeq {32{\cal V}_s^3\over {3\beta^2W_0^2}}a_n^2A_n^2\sqrt{\tau_n}(a_n\tau_n)^2e^{-2a_n\tau_n},\no
&&\eta\simeq {4a_nA_n{\cal V}_s^2\over {3\lambda_n \sqrt{\tau_n}\beta W_0}}\times 4 (a_n\tau_n)^2e^{-a_n\tau_n},\no
&&\xi \simeq {32 (a_nA_n)^2 {\cal V}_s^4\over {9\beta^2 \lambda^2W_0^2\tau_n}}\times 8(a_n\tau_n)^4e^{-2a_n\tau_n}.
\ea
It is required that
\be
{V^{3\over 2} \over {M^3_pV'}}=5.2 \times 10^{-4},\label{52}
\ee
to match the COBE normalization for the density fluctuations $\delta_H=1.92\times 10^{-5}$. We have used the relation $\delta(k)_H^2={1\over {75\pi^2 M^6_P}}{V^3\over V'^2}$.
The left hand side is evaluated at horizon exit, $N_e=50-60 $ e-foldings before the end of inflation.
If we use the relation with $e^{a_n\tau_n}\gg {\cal V}_s^2$
\be
{V^{3\over 2}\over {M_P^3V'}}=\sqrt{V}{1\over \sqrt{\epsilon}}\simeq \sqrt{V_0}{1\over \sqrt{\epsilon}},
\ee
we get the following relation:
\be
{8W_0^2 a^2 N_e^2 \sqrt{\tau_n}\over {A_n^2{\cal V}_s^2}}\simeq 2.7\times 10^{-7}.\label{3.21}
\ee
From Eq. (\ref{3.21}), we can estimate the volume.
If we set $N_e=55$, ${A_n\over W_0}\sim 1$, and $a_n=2\pi$, then this equation is reduced to
\be
{\sqrt{\tau_n}\over {\cal V}_s^2}\simeq 2.83\times 10^{-13}.
\ee
From the analysis of WMAP data, for $47\leq N_e \leq 61$, we have $2.30 \times 10^{-13}\leq \sqrt{\tau_n}/{\cal V}^2_s \leq 3.87\times 10^{-13}$.
We see then that for specific number of e-folding, say $N_e=55$, the volume is roughly given by
\be
{\cal V}_s\simeq 1.88 \tau_n^{1\over 4}\times 10^6.
\ee
By numerical study of Eq. (\ref{ne}), the volume for $47\leq N_e \leq 61$ is roughly $ {\cal V}_s \sim 3.1\times 10^6~ (\tau_n \sim 7.0)$.
\begin{figure}
\begin{center}
\includegraphics[width=2.8in]{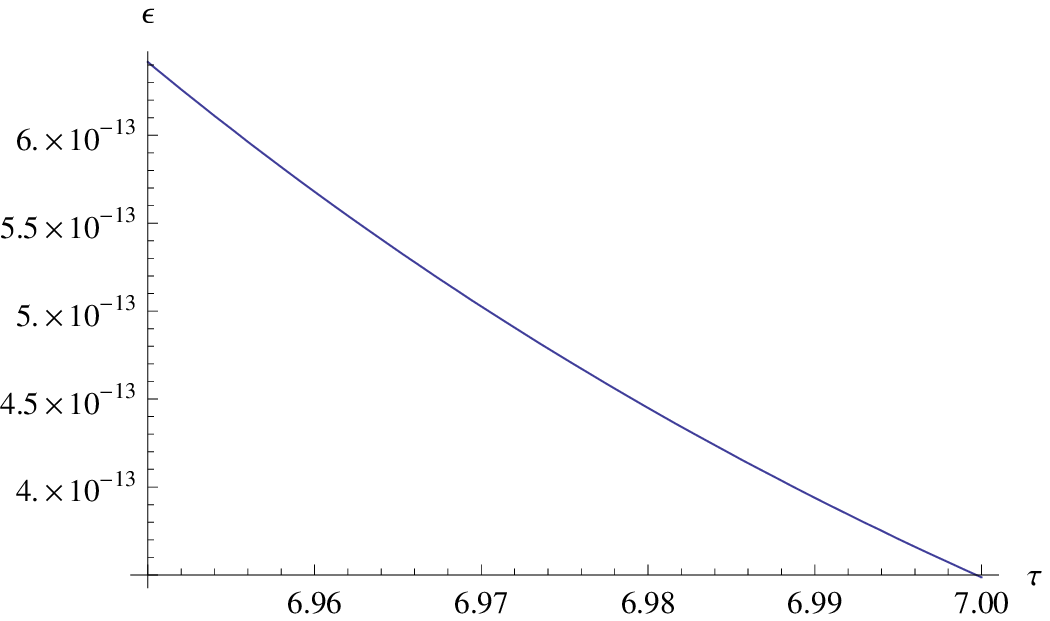}
\includegraphics[width=2.8in]{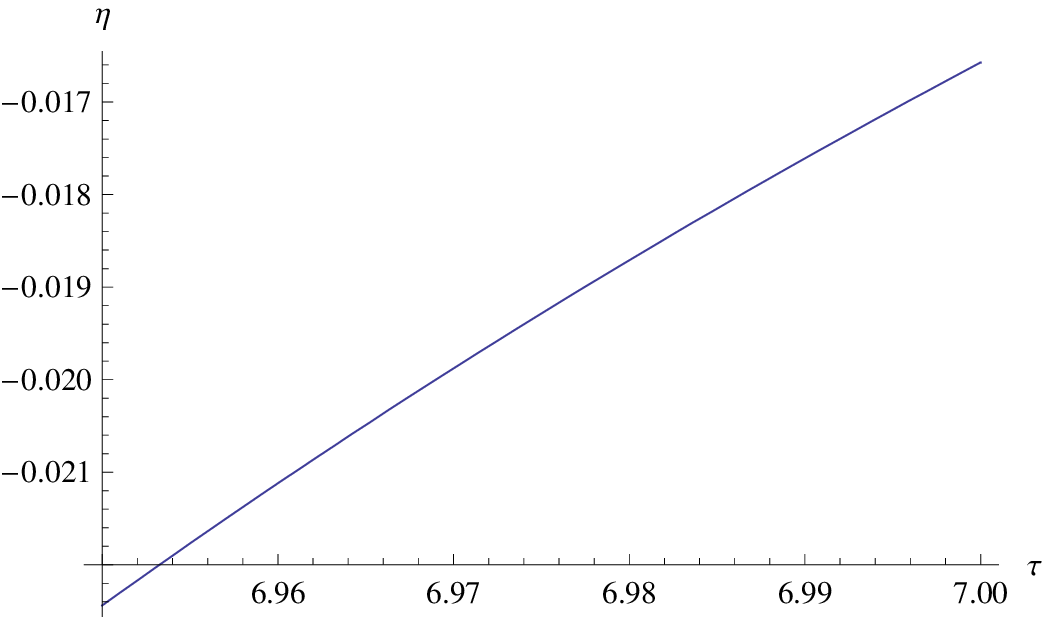}
\includegraphics[width=2.8in]{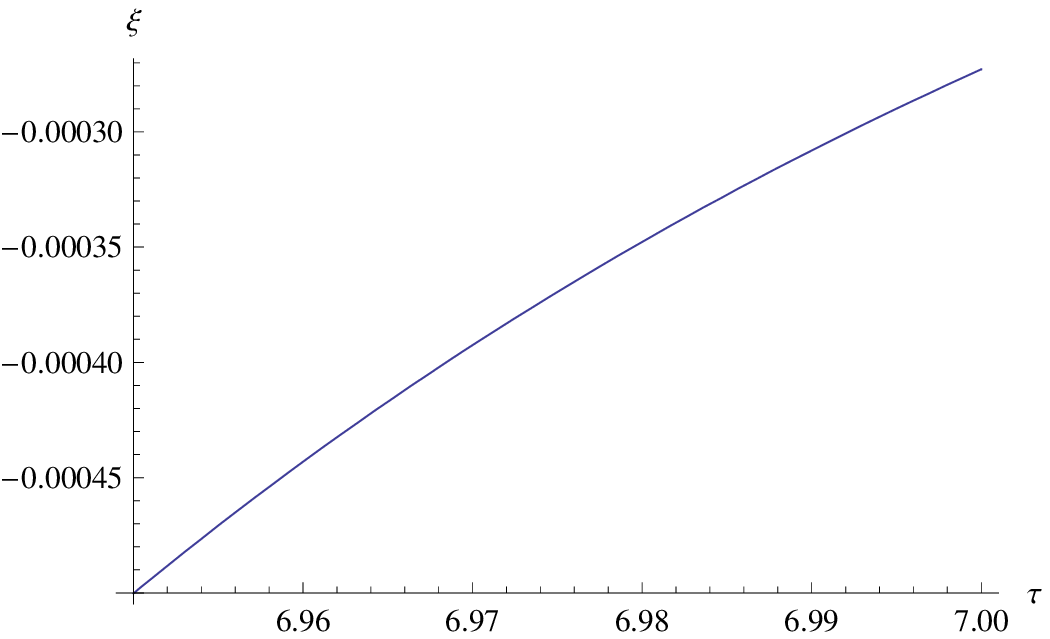}
\end{center}
\caption{Plot of slow roll parameters $\epsilon$, $\eta$ and $\xi$, when $N_e=55$ with numerical values ${\cal V}_s=3\times 10^6$, $\hat W_0=1, \beta=1, a_n=2\pi, \hat A_n=1$, and $\lambda=1$. }\label{k6}
\end{figure}
With the help of the expression of $N_e$ and $\epsilon$ we can see
\be
\epsilon= {\lambda^2 \over {a_n^3 \sqrt{\tau_n}N_e^2{\cal V}_s}}=1.33\times 10^{-13},~~{\rm for}~~{\cal V}_s=3.1 \times 10^6,~\lambda=1,\tau_n=7.0, N_e=61.
\ee
When we consider $\lambda \simeq {\cal O}(1)$ and with the above relations, the rough estimate for slow roll parameters $\epsilon$, $\eta$ and $\xi$ become
\be
\epsilon < 10^{-12},~~~\eta \simeq -{1\over N_e},~~~\xi \simeq {1\over N^2_e}.
\ee
This estimation is checked explicitly by numerical evaluation in figure (\ref{k6}). Note that these values of $\eta$ and $\xi$ are independent of the parameters of the theory (e.g. $\alpha$, ${\cal V}_s$, and $ W_0$, etc.) and are dependent only on the number of e-folding. On the other hand, $\epsilon$ is dependent on the particular value of ${\cal V}_s$ and $W_0$, etc. Later on, in the model with the number of K\"{a}hler moduli more than two, we will find similar behavior for the slow roll parameters $\eta$ and $\xi$. Finally, we plot the spectral index $n_s$ as a function of $\tau$ in figure (\ref{k123}).

\section{Examples of the Swiss-cheese model}
\setcounter{equation}{0}
\subsection{Two K\"{a}hler moduli model: $P^4_{1,1,1,6,9}$ }
In this section, we consider concrete Calabi-Yau manifolds characterized by weighted projective space called Swiss-cheese model. One of the advantages using this model is that we can explicitly express the volume of the Calabi-Yau manifold and give a chance to calculate many things. A Swiss-cheese model is a real 6-manifold with 4-cycles. Of these 4-cycles, one $\tau_b$ is large and controls the size of the cheese, while the others $\tau_{s,i}$ are small and controls the size of the holes. The volume of the cheese can be expressed as in Eq. (\ref{volume}).
The small cycles may be thought of as local blow-up effects; if the bulk cycle is large, the overall volume is largely insensitive to the small cycles.

In this subsection, the case is the example of Calabi-Yau of degree 18 hypersurface embedded in the complex weighted projective space $P^4_{1,1,1,6,9}$ \cite{balasubb,balasuba}. The overall volume in terms of 2-cycle volumes $t$'s is given by
\be
{\cal V}_s={1\over 6} (3 t^2_1t_5+18t_1t^2_5+36t^3_5).
\ee
With the four-cycle volumes such as $\tau_4={t^2_1\over 2}$, $\tau_5={(t_1+6t_5)^2\over 2}$, the of volume Calabi-Yau manifold takes the form
\be
{\cal V}_s={1\over {9\sqrt{2}}}\(\tau^{3\over 2}_5-\tau^{3\over 2}_4\).
\ee
The large volume claims that $\tau_5\to \infty$ and $\tau_4$ remains small.
The superpotential is then given by
\be
W=W_0+A_4e^{-a_4\tau_4}.
\ee
We take the limit ${\cal V}_s\to\infty (\tau_5\to \infty)$ with $a_4\tau_4=\log {\cal V}_s $.
Then, the potential becomes
\be
V\sim {1\over {\cal V}_s}a_4^2|\hat A_4|^2\sqrt{\tau_4}e^{-2a_4\tau_4}-{1\over {\cal V}_s^2}a_4\tau_4e^{-a_4\tau_4}|\hat A_4\hat W_0|+{\xi\over{{\cal V}_s^3}}|\hat W_0|^2.
\ee

In order to discuss the stabilization we have to solve
\be
{\partial V\over {\partial {\cal V}_s}}={\partial V\over {\partial \tau_4}}=0,
\ee
and by taking the limit $a_4\tau_4\gg 1$ we get the following:
\be
\tau_4\sim (4\xi)^{2\over 3},~~~{\rm and}~~~{\cal V}_s \sim {\xi^{1\over 3}|W_0|\over {a_4 A_4}}e^{a_4\tau_4}.
\ee
So we have seen the stabilization of the potential with the given $\tau$ and the volume.
For the discussion of the numerical analysis, we have
\be
\xi=1.31.
\ee
If we set $\hat A_4=1$, $a_4=2\pi$ and $\hat W_0=1$, we have
\be
{\cal V}_s\approx 2.84 \times 10^7.
\ee

Finally, let us consider inflationary potential. Following the procedure of the previous section, we also get the same form of the potential:
\be
V={\xi\over{{\cal V}_s^3}}|\hat W_0|^2-{1\over {\cal V}_s^2}a_4\tau_4e^{-a_4\tau_4}|\hat A_4\hat W_0|.
\ee

\subsection{The three K\"{a}hler moduli model: ${\cal F}_{11} $ and $P^4_{1,3,3,3,5}$}
\subsubsection{Case I: $ {\cal F}_{11} $ }
In the case of three K\"{a}hler moduli, the potentials were calculated \cite{denefdoug}. In this section, we get the inflationary potential by the same procedure with section 3.

We start with an example, the Fano three-fold ${\cal F}_{11}$ \cite{denefdoug} which is a $Z_2$ quotiont
of a real six-dimensional Calabi-Yau manifold with hodge numbers $h_{1,1}=3$ and $h_{2,1}=111$.
Then, the volume can be expressed in terms of 4-cycle moduli $\tau_i$ as
\be
{\cal V}_s={1\over {3\sqrt{2}}}\(2(\tau_1+\tau_2+2\tau_3)^{3\over 2}-(\tau_2+2\tau_3)^{3\over 2}
-\tau_2^{3\over 2}\).
\ee
With the diagonal basis
\be
\tau_a=\tau_1+\tau_2+2\tau_3,~~~\tau_b=\tau_2+2\tau_3,~~~\tau_c=\tau_3,
\ee
we can rewrite the volume simpler as
\be
{\cal V}_s={1\over {3\sqrt{2}}}\(2\tau_a^{3\over 2}-\tau_b^{3\over 2}
-\tau_c^{3\over 2}\).
\ee
There is another type of Calabi-Yau manifold which has the same number of K\"{a}hler moduli but different
number of complex structure. We study it in more detail in following section.

\subsubsection{Case II: $P^4_{1,3,3,3,5}$}

This Calabi-Yau manifold has $h_{1,1}=3$ and $h_{1,2}=75$ and has been studied in \cite{blum}. For this case, we have two stacks of D7-branes wrapping rigid four-cycles $D_A$ and $D_B$ where on the first one a non-trivial line bundle ${\cal L}_A$ is turned on. We get MSSM matter from intersections $AA'$ and $AB$ where the prime denotes the orientifold image. The stacks of D7-branes wrap the rigid four-cycles $D_{D7A}=D_5+D_6={1\over 3}(D_b-2D_c)$ and $D_{D7B}=D_5=D_c$
with line bundles ${\cal L}_A={1\over 3}(2D_b+5D_c)$ and ${\cal L}_B=0$.
With this choice, there are no chiral zero modes on the D7-E3 brane intersections.
The E3-brane wraps $D_{E3}={1\over 3}(D_b+D_c)=2D_5+D_6$.
For this case, the volume in terms of $\tau$'s is given by
\be
{\cal V}_s=\sqrt{ {2\over 45}}\[ \(5\tau_1+3\tau_2+\tau_3\)^{3\over 2}-{1\over 3}\(5\tau_1+3\tau_2\)^{3\over 2}-{\sqrt{5}\over 3}\tau^{3\over 2}_1\].
\ee
Again in terms of diagonal basis
\be
\tau_a=5\tau_1+3\tau_2+\tau_3,~~~\tau_b=5\tau_1+3\tau_2,~~~\tau_c=\tau_1,
\ee
we have the volume:
\be
{\cal V}_s=\sqrt{{2 \over 45}}\(\tau^{3\over 2}_a-{1\over 3}\tau^{3\over 2}_b-{\sqrt{5}\over 3}\tau^{3\over 2}_c\).
\ee
For the model $P^4_{1,3,3,3,5}$, \cite{blum} we write the scalar potential as
\be
V={\lambda_1(\sqrt{5\tau_b}+\sqrt{\tau_c})e^{- {4\pi \over 3}(\tau_b+\tau_c)}\over {\cal V}_s}
-{\lambda_2(\tau_b+\tau_c)e^{-{2\pi\over 3}(\tau_b+\tau_c)}\over {\cal V}_s^2}+{\lambda_3\over {\cal V}_s^3}.
\ee
where $\lambda_i>0, i=1,2,3$ are numerical factors. Changing of coordinates, using Euclidean D3-brane cycle $\tau_{E3}$ and standard model cycle $\tau_{SM}$ with the relations $\tau_b=2\tau_{E3}+\tau_{SM}$ and
$\tau_c=\tau_{E3}-\tau_{SM}$, gives the potential:
\be
V={\lambda_1\(\sqrt{5(2\tau_{E3}+\tau_{SM})}+\sqrt{\tau_{E3}-\tau_{SM}}\)e^{-4\pi \tau_{E3}}\over {\cal V}_s}-{3\lambda_2\tau_{E3}e^{-2\pi \tau_{E3}}\over {\cal V}_s^2}+{\lambda_3\over {\cal V}_s^3}.
\ee

From this potential, we can get inflationary potential:
\be
V={\lambda_3\over {\cal V}_s^3}-{3\lambda_2\tau_{E3}e^{-2\pi \tau_{E3}}\over {\cal V}_s^2},
\ee
which has the same form with the previous potentials.

\begin{figure}
\begin{center}
\includegraphics[width=2.8in]{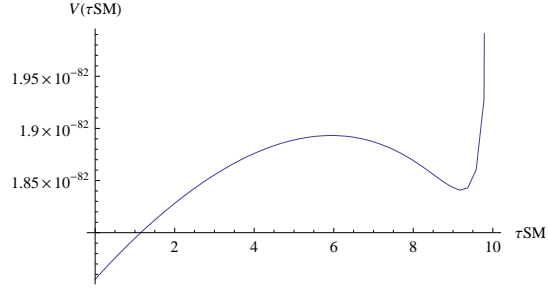}
\end{center}
\caption{Here we take $l_1=1$, $l_2=1$, $l_3=1$, $l_4=5$ and $l_5=5$.  }\label{F1}
\end{figure}

The potential has a critical point at $\tau_{E3}=2\tau_{SM}$ but this is not a minimum but a saddle point along $\tau_{SM}$ at fixed $\tau_{E3}$ and ${\cal V}_s$.
The inclusion of the one-loop $g_s$ correction $\delta V^{KK}_{(g_s)}$ stabilizes this direction.
So the one loop corrected scalar potential \cite{cicob} has the form
\ba
V_F&=&{\lambda_1\(\sqrt{5(2\tau_{E3}+\tau_{SM})}+\sqrt{\tau_{E3}-\tau_{SM}}\)e^{-4\pi \tau_{E3}}\over {\cal V}_s}
-{3\lambda_2\tau_{E3}e^{-2\pi \tau_{E3}}\over {\cal V}_s^2}+{\lambda_3\over {\cal V}_s^3}\no
&+&\( {5\over {\sqrt{\tau_{E3}-\tau_{SM}}}}+{13\sqrt{5}\over {\sqrt{2\tau_{E3}+\tau_{SM}}}}\) {1\over {15\sqrt{2}{\cal V}_s^3}}.
\ea
Figure (\ref{F1}) depicts the stabilization of the direction $\tau_{SM}$.

The model with D7-brane induces D-term. The D-term contribution to the potential becomes\cite{blum}
\be
V_D={\lambda_4\over {\cal V}_s^2}{1\over {\tau_b-\tau_c}}\(\sqrt{5\tau_b}-\sqrt{\tau_c}\)^2.
\ee
Note that because of D-term the large volume minimum is destabilized: For the large volume scenario, we need to set $\tau_b=5\tau_c$. Now, let us consider the inflation with this potential. As we have seen, the inflationary potential is obtained by neglecting the ${1\over {\cal V}_s}$ term. Moreover, we set $\tau_b=2\tau_{E3}+\tau_{SM}$ and $\tau_c=\tau_{E3}-\tau_{SM}$. The inflaton is assumed to be $\tau_{E3}$ and this scalar field is supposed to be rolling from off the minimum. Then D-term plays a role in this case. However, if we compute the slow parameters $\epsilon$ and $\eta$ with the inclusion of D-terms the potential is not flat. D-term dominates in the potential and it behaves linearly in $\tau_{E3}$.

\subsection{Four moduli K\"{a}hler model: ${\bf P}^4_{1,2,2,10,15}$}

In order to accommodate two MSSM-like D7-stacks and E3-brane, all on different `small' cycles, we need Calabi-Yau manifolds with at least four moduli. Since they will generically intersect the E3-branes, the E3-D7 strings will correspond to chiral zero-modes of the instanton. We have E3-brane placed on a small four-cycle and two MSSM-like D7-branes with unitary gauge groups placed on the two remaining four-cycles. In order to cancel the total D7 tadpole by the O7-plane, a hidden D7-brane is needed. Here, three out of four moduli come from blow-ups. There are four Swiss-cheese models, and we focus here on just one model: ${\bf P}^4_{1,2,2,10,15}$ \cite{coll}.
The volume can be expressed as
\be
{\cal V}_s ={ \sqrt{2}\over 3} \[ {1\over {7\sqrt{3}}} \( 15\tau_1 +9\tau_2 +7\tau_3+3\tau_4\)^{3\over 2}-{1\over 35}\(5\tau_1+3\tau_2+\tau_4\)^{3\over 2}
-{1\over 15}\(3\tau_2+\tau_4\)^{3\over 2}-{1\over 3}\tau_4^{3\over 2} \].
\ee
For positive K\"{a}hler cone $\int J >0$, the following conditions should be satisfied:
\be
\tau_2,\tau_3\to \infty,~~\tau_1,\tau_4\to 0,~~{\rm or}~~\tau_1,\tau_3\to \infty,~~\tau_2,\tau_4\to 0.
\ee
We choose the diagonal basis:
\ba
\tau_a&=&15t_1+9t_2+7t_3+3t_4,\no
\tau_b&=&5t_1+3t_2+t_4,\no
\tau_c&=&3t_2+t_4,\no
\tau_d&=&t_4.
\ea
In this basis, the volume can be rewritten as
\be
{\cal V}_s={\sqrt{2} \over 3} \( {1\over {7\sqrt{3}}}\tau^{3\over 2}_a-{1\over 35}\tau^{3\over 2}_b-{1\over 15}\tau^{3\over 2}_c-{1\over 3}\tau^{3\over 2}_d\).
\ee
We can divide our model into two scenarios. It depends on which divisor $E3$, $D7_A$ and $D7_B$ branes are placed. For scenario I, $E3$, $D7_A$ and $D7_B$ are placed on $\eta_1$, $\eta_4$, and $\eta_2$, respectively. While for scenario II, $E3$, $D7_A$ and $D7_B$ are placed $\eta_4$, $\eta_1$, and $\eta_2$, respectively.

The $F$-term potential for scenario I is given by
\be
V_F={1\over { {\cal V}_s}} {4\pi^2\sqrt{2}\over 5} (7\sqrt{5{ \tau}_{E3}+\tau_c}+3\sqrt{\tau_c})|\hat A_{E3}|^2e^{-4\pi {\tau}_{E3}}-{1\over {{\cal V }_s}^2}
4\pi{\tau}_{E3}e^{-2\pi {\tau}_{E3}}|\hat A_{E3}\hat W_0|+{{3\xi}\over {4{\cal V}_s}^3}|\hat W_0|^2.
\ee
By varying with ${{\cal  V}}$, $\partial V_F/\partial {{\cal  V}}=0$, we get
\be
{ {\cal V}_s}=e^{2\pi \tau_{E3}}\tau_{E3} {{ \( 20\pm \sqrt{1600-45\(3\sqrt{2\tau_c}+7\sqrt{2(\tau_c+\tau_{E3}) } \) } \)} \over { 4A_{E3}\pi (\sqrt{2\tau_c}+7\sqrt{\tau_c+5\tau_{E3}})  }  }.
\ee
Note that if we demand a large volume while at the same time fulfilling the K\"{a}hler cone constraints, the term in the square root becomes negative. We cannot realize the large volume in scenario I. Therefore, it is inappropriate to discuss the stabilization of this model.

For scenario II, we see that the potential is written as
\be
V_F={1\over {{\cal V}_s}} 12\pi^2 \sqrt{2{\tau}_{E3}}|\hat A_{E3}|^2e^{-4\pi {\tau}_{E3}}-{1\over {{\cal V}_s}^2}4\pi {\tau}_{E3}e^{-2\pi {\tau}_{E3}}|\hat A_{E3}\hat W_0| +{{3\xi} \over {4{\cal V}_s}^3}|\hat W_0|^2.
\ee
This potential is also similar to scenario I whereas it has a large volume limit. We can proceed the same procedure to find the inflationary potential:
\be
V={{3\xi} \over {4{\cal V}_s}^3}|\hat W_0|^2-{1\over {{\cal V}_s}^2}4\pi {\tau}_{E3}e^{-2\pi {\tau}_{E3}}|\hat A_{E3}\hat W_0|.
\ee

It is unpredicted that the potentials in the large volume scenario for different number of K\"{a}hler modulus give the same type of inflationary potential. After diagonalizing the volumes of the Calabi-Yau of the weighted projective spaces, the volumes can be rewritten as ${\rm Vol_L}$-${\rm Vol_S}$. Within this type of volume, the potentials in the large volume limit effectively are the same with just two K\"{a}hler moduli. In the next section, we study two types of potential as an inflation including the one above.

Now, let us relate the above potentials to inflationary potentials. So far, we have omitted the discussion about the kinetic term of the moduli. For the potentials we have, the kinetic term would be of the form ${1\over 2} K_{ij}\partial_\mu \tau_i\partial_\nu \tau_j$. In ordinary inflationary model, the kinetic term is of the form ${1\over 2}(\partial \phi)^2$. Hence, by absorbing the K\"{a}hler metric into the moduli $\tau$, we get such a kinetic term with canonically normalized field, $\phi$. In doing so, we arrive at the form of the potential $V=V_0-\alpha\phi^{4\over 3} e^{-\phi^{4\over 3}}$.
In the next section, we will study the behavior of this type of potential and will try to see how the WMAP data can be fit. We can ask whether the power of the field, ${4\over 3}$, is important. It turns out that it is so. In order to demonstrate that the theory is very sensitive to the power of the inflaton field in the potential, we first consider a toy model $V=V_0(1-\alpha\phi e^{-\phi})$.
\section{Inflations with generic potentials}
\setcounter{equation}{0}
In this section, we study two inflationary potentials. The first one is the potential of the toy model which has a form close to the potential of the K\"{a}hler moduli above. The second one is the potential of the K\"{a}hler moduli which is discussed in section 3 but with canonically normalized inflaton field. We test these potentials whether these two potentials are good candidates as inflationary models comparing with the recently released seven-year WMAP data.

\subsection{ Case I: A toy model $V=V_0(1-\alpha \phi e^{-\phi})$}
We have seen a few examples of the inflationary potentials with the number of K\"{a}hler moduli,
two, three, or four. They all can be reduced to Swiss-cheese models at least by diagonalization procedure. If we take the large volume limit for these models we get the same form of potential. As inflationary models, they all have the same type of K\"{a}hler potential such as $V=V_0(1-\alpha \phi^{4\over 3}e^{-\phi^{4\over 3}})$.

Recently, the seven-year WMAP data were released \cite{7wmap}. For $\Lambda$CDM model, the seven-year WMAP data only show that the index of the power spectrum satisfies $n_s=0.963\pm 0.014$.
Combining WMAP data with BAO and $H_0$, the result is $n_s=0.963\pm 0.012$ which is comparable to the earlier data.

The running of the spectral index with seven year data only is given by
\be
\alpha_s={dn_s\over {d\ln k}}=-0.034\pm 0.021,
\ee
while combining with BAO and $H_0$ we have
\be
\alpha_s={dn_s\over {d\ln k}}=-0.022\pm 0.020.
\ee
Note that the result of the running of the spectral index is a little bit bigger than the result of the five-year data.

In Ref.\cite{liddlyth2}, it has been proposed that the number of e-folding relevant to the observation should be $N_e=54\pm 7$.

Before studying the string theoretic inflationary potential, in this section, we first study a toy model with the simpler form. That is, we start with the potential given by
\be
V=V_0(1-\alpha\phi e^{-\phi}),
\ee
which plays a role model for the potentials studied in the previous sections. If we forget the kinetic term and set $\phi^{4\over 3}$ to $\phi$ in the potential, the potential is reduced to this form. In order to analyze the cosmological predictions we need to find the slow roll parameters.
First, with this potential and its derivatives, now we can calculate the number of e-folds $N_e$:
\be
N_e\simeq {1\over \alpha}{e^{\phi}\over {\phi}},~~{\rm and}~~\phi\gg 1. \label{eqne1}
\ee
To simplify the analysis, we take the limit $\phi \gg 1$. Note that in the final form there is no $\alpha$ dependence. The dependence comes only at subleading order.

For a given number of e-folding $N_e$, we can solve the above equation Eq.(\ref{eqne1}) to find $\phi$ and we get the following solution:
\be
\phi=1-W_n\(-{e\over {\alpha N_e}}\),~~~\alpha N_e\neq 0,~n\in Z
\ee
where $W_k(z)$ is the analytic continuation of the product log function and is called Lambert function\cite{lambert}. The integer $k$ denote the branch of $W_k(z)$. $W_0(z)$ and $W_{-1}(z)$ are the only branches of the Lambert function. The Lambert function is the inverse function of $f(w)=we^w$. 
One of the properties of Lambert function is the following:
\be
W_0(0)=0,~~W_0\(-{1\over e}\)=-1,~~\lim_{x\to 0^-} W_{-1}(x)=-\infty.
\ee
The Taylor expansion of $W_0(x)$ is
\be
W_0(x)=\sum^{\infty}_{n=1} { (-1)^{n-1}\over {n!}}=x-x^2+{3\over 2}x^3\cdots.
\ee

The slow roll parameters $\epsilon$, $\eta$, and $\xi$ can be found and expressed as follows:
\be
\epsilon \simeq {1\over {2{N_e}^2}},~~\eta \simeq -{1\over N_e},~~\xi \simeq {1\over N_e^2}.
\ee
In these expressions, we used the relation between the number of e-folds $N_e$ and the inflaton field $\phi$ in Eq.({\ref{eqne1}).

Now let us consider the spectral index $n_s$ \cite{liddlyth}. Its physical meaning is that the fluctuation generated by slow roll inflation is given by \cite{lidd}
\be
P_s(k)\sim k^{n_s-1}.
\ee
Hence, when we consider the case $n_s=1$, then the fluctuation is independent of $k$.
The scalar to tensor ratio $r$ is given by $r=16\epsilon$. So for the range of the number of e-folding $47<N_e<61$, we have $2.10\times 10^{-3}<r<3.60\times 10^{-3}$.
Now let us compute the spectral index and its running \cite{turner}. They are given by
\ba
n_s&\simeq &1+2\eta-6\epsilon\simeq 1-{2 \over N_e}-{3\over { N_e}^2},\no
\alpha_s&=&{dn_s\over {d\ln k}}\simeq 16\epsilon\eta-24\epsilon^2-2\xi\simeq -{2\over N_e^2}-{8\over { N_e^3}}-{6\over { N_e^4}}.
\ea
With the given two equations and we can express the running $\alpha_s$ as a function of $n_s$:
\be
\alpha_s=-{2\over 27}\(4+2\sqrt{4-3n_s}-15n_s+9n_s^2\).\label{nsalphas}
\ee
We plot $\alpha_s$ in terms of $n_s$ in figure (\ref{alphavns1}).
\begin{figure}
\begin{center}
\includegraphics[width=2.8in]{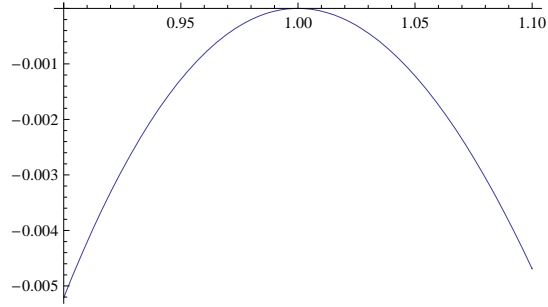}
\end{center}
\caption{The running $\alpha_s$ versus $n_s$.}\label{alphavns1}
\end{figure}
For the e-folding $47<N_e<61$, the spectral index $n_s$ and the running $\alpha_s$ are given by
\ba
0.956<&n_s&<0.966,\no
-9.84\times 10^{-4}<&\alpha_s&<-5.73\times 10^{-4}.
\ea
When we compare this with the seven-year data shown, $n_s$ fit the seven-year WMAP date while $\alpha_s$ is too small and does not fit the seven-year WMAP data.

Conversely, in order to fit the seven-year WMAP data $0.949<n_s<0.977$ or $0.951<n_s<0.975 $ (WMAP+BAO+$H_0$), the range of the number of e-folding should be
\be
40.7<N_e<88.4~~~~{\rm or}~~~~42.3<N_e<81.5.\label{range}
\ee
This range is wide enough to cover whole part of the range $47<N_e<61$.
Furthermore, for the given ranges of the number of e-folding, we can estimate the range of $\alpha_s$:
\be
1.33\times 10^{-3}<\alpha_s<2.67\times 10^{-4} ~~~~{\rm or}~~~~1.23\times 10^{-3}<\alpha_s<3.08\times 10^{-4}.
\ee
Of course, we see that this range of $\alpha_s$ is outside of the observational seven year WMAP data.
In summary, our toy model gives a quite reasonable spectral index while the running deviates from the observational data.
\subsection{Case II: $V=V_0(1-\alpha \phi^{4\over 3}e^{-\phi^{4\over 3}})$}
We have studied a few examples with K\"{a}hler moduli inflationary potentials with the number of moduli two, three, or four within the Swiss-cheese models. As a result, we have seen that they all give the same type of potentials while they have different number of flat directions. Nevertheless, it seems at this moment that the potential of this type $V=V_0(1-\alpha \phi^{4\over 3}e^{-\phi^{4\over 3}})$ is generic for Swiss-cheese models. In this section, we investigate the potential comparing with the seven-year WMAP data.

We considered the potential in section 3:
\be
V=V_0 -{4\hat W_0a_n\hat A_n \over {\cal V}_s^2}\({3{\cal V}_s\over {4\lambda}}\)^{2\over 3} \phi^{4\over 3}e^{-a_n ({3{\cal V}_s\over {4\lambda}})^{2\over 3}\phi^{4\over 3}},
\ee
where we have the relation between the normalized inflaton field, $\phi$, and the four-cycle, $\tau$,
\be
\phi=\sqrt{{4\lambda \over {3\cal V}_s}}\tau^{3\over 4}.
\ee
In order to handle easily, we rewrite this potential as a simple form:
\be
V=V_0-a \phi^{4\over 3}e^{-b\phi^{4\over 3}},~~b=a_n\({3{\cal V}_s\over {4\lambda}}\)^{2\over 3},~~a=b{4\hat W_0\hat A_n\over {\cal V}_s^2},~~~V_0={\beta \hat W_0^2\over {\cal V}_s^3}.
\ee
We calculate the number of e-folding $N_e$ for the limit $e^\phi \gg 1$:
\be
N_e\simeq {9V_0\over {16ab^2}}{e^{b\phi^{4\over 3}}\over \phi^2}.\label{eqne2}
\ee
We can also find $\phi$ by solving the above equation:
\ba
&&\phi^2 N_e =C_0 e^{b\phi^{4\over 3}},\no
&&\phi=\pm  \(-{3\over 2}\)^{3\over 4}b^{-3\over 4}W_0\(-{2bC_0^{2\over 3}\over {3N_e^{2\over 3}}}\)^{3\over 4}.
\ea
Then, with the help of the expression of the number of e-folding $N_e$ in Eq.(\ref{eqne2}), the slow roll parameters are given by \ba
\epsilon&=&{1\over 2}M_p^2 \({V'\over V}\)^2 \simeq{1\over 2}\({4ab\over {3 V_0}}\)^2\({\phi^{5\over 3}\over e^{b\phi^{4\over 3}}}\)^2,\no
\eta &=& M_p^2{V''\over V}\simeq - {1\over N_e},\no
\xi &=& M_p^4{V'V'''\over V^2}\simeq {1\over N_e^2}.\label{srp}
\ea
For the expressions of both $\eta$ and $\xi$, we obtain the same formula with the one obtained in section 3. For $\epsilon$, as we have seen in section 3, we get the relation $\epsilon < 10^{-12}$. As was done in Eq.(\ref{nsalphas}), by combining the slow roll parameters in Eq.(\ref{srp}), we can find the $\alpha_s$ as a function of $n_s$:
\be
\alpha_s=-{1\over 2}(1-n_s)^2.
\ee
In figure (\ref{alphasvns2}), we plot $\alpha_s$ in terms of $n_s$.

\begin{figure}
\begin{center}
\includegraphics[width=2.8in]{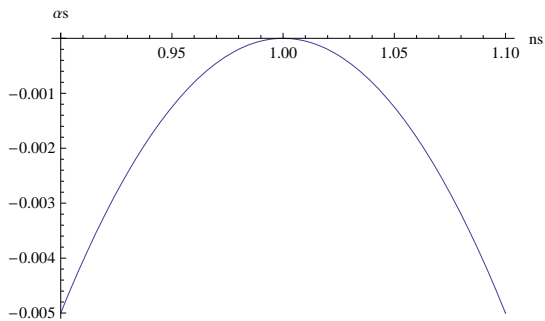}
\end{center}
\caption{The running $\alpha_s$ versus $n_s$.}\label{alphasvns2}
\end{figure}

Now, let us consider the spectral index and its running with the number of e-folding predicted by WMAP data. Taking a range of $47<N_e<61$, we obtain $n_s$ and $\alpha_s$:
\ba
0.96&<&n_s<0.97,\no
-9.10\times 10^{-4}&<&\alpha_s<-5.37\times 10^{-4}.
\ea
Just as we have seen in the result for the previous toy model, the range of spectral index fit the seven-year WMAP data while its running $\alpha_s$ does not fit the seven-year WMAP data.
In order to get the same range of the running $\alpha_s$ with the seven-year WMAP data, we require that the range of the number of e-folding be $6.03<N_e<12.40$ or $6.90<N_e<10.00$. It is too small for the required number of e-folding. Since $\epsilon$ is too small to be neglected, we drop the dependence of the term $\epsilon$ in the numerical calculation. So, the tensor to scalar ratio $r=16\epsilon$ is also negligibly small.

In this subsection, we have been used the normalization of the analysis of COBE data studied in \cite{bunn}. For the three-year WMAP, we have the normalization \cite{liddle}
\be
\delta_H^{WMAP3}=1.93\times 10^{-5} \sqrt{Q},
\ee
where $Q$ is given by
\be
Q={e^{-2(1.24+1.04r)(1-n_s)}\over {1+0.53r}}.
\ee
From this result, we get the following:
\be
{V^{3\over 2}\over {M_p^3 V'}}=5.25\times 10^{-4}\sqrt{Q}.
\ee
So, for $n_s=1$ and $r=0$, we have $Q=1$, and therefore we have $\delta_H^{WMAP3}=1.93 \times 10^{-5}$,
giving ${V^{3\over 2}\over {M_p^3 V'}}=5.25\times 10^{-4}$.

With the help of Ref.\cite{liddle}, Eq.(\ref{3.21}) can be expressed as
\be
{32 \pi^2 N_e^2\sqrt{\tau}\over {\cal V}_s^2}\simeq\(2.8\times 10^{-7}\)Q.
\ee
Let us consider the number of e-folding as $N_e=55$. Then we have
\be
{\sqrt{\tau}\over {\cal V}_s^2}\simeq 2.93\times 10^{-13}Q.\label{555}
\ee
Hence, we have ${\cal V}_s\simeq 1.84 \times 10^6 \sqrt{Q}$.
For the range of the number of e-folding $47<N_e<61$, we have $2.38\times 10^{-13}<Q<{\sqrt{\tau}\over {\cal V}_s^2}<4.01\times 10^{-13}Q$. In section 3, we saw that the volume ${\cal V}_s$ is roughly $3.1\times 10^6$ when the moduli $\tau$ is roughly 7. With these volume ${\cal V}_s$ and the moduli $\tau$, we get ${\cal V}_s^2\simeq 2.75 \times 10^{-13}$. We can guess the value of $Q$ from Eq.(\ref{555}): the range of $Q$ should be $0.68<Q<1.16$. Since the maximum of $Q$ is 1, then we have $0.68<Q<1$. The value of the spectral index $n_s$ giving $Q=0.68$ is $0.84$.

When we have $Q=1$, this equation becomes the same with the one coming from the analysis of COBE data. Let us consider the value of $Q$ for the spectral index of the seven-year WMAP data. When the spectral index $n_s$ and the tensor to scalar ratio $r$ have values $n_s=0.96$ and $r=0$ respectively, we have $Q\simeq 0.91$. When we have $n_s=0.97$ and $r=0$, we have $Q\simeq 0.93$. For non-zero $n_s$, the factor $Q$ is less than zero. Because we have $Q<1$, the right hand side becomes smaller than the case of COBE normalization.

\section{Conclusion and discussion}
In this paper, we considered the inflationary models whose potentials come from type IIB string theory compactified on Calabi-Yau manifolds with fluxes. The complex structure moduli and axionic field are fixed by fluxes while the K\"{a}hler moduli are fixed by the non-perturbative superpotential. We investigated the particular weighted projective spaces with different number of K\"{a}hler moduli. In these models, the biggest four-cycle of the Calabi-Yau manifolds is fixed and controls the bulk Calabi-Yau volume. One of the rest of four-cycles, the smallest one, plays the role of the inflaton field. Using the large volume limit, the potentials for the smallest four-cycle cycle were given. The inflationary potential can be derived \cite{conquev} by taking the special limit for the four-cycle. In the present work, we introduced models with two, three, or four K\"{a}hler moduli. We found that all three cases give the same form of inflationary potential.

As a concrete example, we studied two inflationary models. With the given potentials, we could calculate the number of e-folding, $N_e$, and slow roll parameters, $\epsilon$, $\eta$, and $\xi$. Then, the spectral index, $n_s$, and its running, $\alpha_s$ were obtained from these information.
We first investigated a toy model which has no string theoretic origin. The parameters $\epsilon$, $\eta$, and $\xi$ have magnitudes on the order of $1/2N_e^2$, $-1/N_e$, and $1/N_e^2$, respectively. At leading order, these parameters are independent of the parameter of the potential. In the second model, we reanalyze the K\"{a}hler moduli inflation in terms of canonically normalized inflaton field. Just as we have seen in section 3, the parameters $\epsilon$, $\eta$ and $\xi$ have magnitudes $<10^{-12}$, $-1/N_e$ and $1/N_e^2$, respectively. The slow roll parameters of this model are also free from the parameter of the potential at leading order. Then, we computed the $n_s$ and $\alpha_s$ for each model. In order to see the viability of both models, we compared the spectral index and its running with the recently released seven-year WMAP data for each model. The spectral index $n_s$ has reasonable value while the running $\alpha_s$ has the same sign but smaller than the seven-year WMAP.
Conversely, we calculated the range of the number of e-folding for the given ranges of $n_s$ of the seven-year WMAP data. The first model gave reasonable range while the second model gave the range of smaller value than the required number of e-folding.
\section{Acknowledgements}
This work was supported by the National Research Foundation of
Korea Grant funded by the Korean Government [NRF-2009-351-C00018].

\end{document}